\documentclass[twocolumn,superscriptaddress]{revtex4-1}%
\usepackage{amsmath}%
\usepackage{amsfonts}%
\usepackage{amssymb}%

\usepackage{graphicx}

\usepackage{fullpage}

\usepackage{dsfont}
\usepackage{hyperref}
\usepackage{bm}
\usepackage{multirow}
\usepackage{color}
\usepackage{float}
\usepackage{hyperref}
\usepackage{xcolor}

\begin{document}

\title{Impact of charge transfer excitons on unidirectional exciton transport \\ in lateral TMD heterostructures}

\author{Roberto Rosati}
\email{roberto.rosati@physik.uni-marburg.de}
\affiliation{Department of Physics, Philipps-Universit\"at Marburg, Renthof 7, D-35032 Marburg, Germany}

\author{Sai Shradha}
\affiliation{Institute of Condensed Matter Physics, Technische Universität Darmstadt, 64289 Darmstadt, Germany}

\author{Julian Picker}
\affiliation{Institute of Physical Chemistry, Friedrich Schiller University Jena, 07743 Jena, Germany}
\affiliation{Abbe Centre of Photonics, 07745 Jena, Germany}

\author{Andrey Turchanin}
\affiliation{Institute of Physical Chemistry, Friedrich Schiller University Jena, 07743 Jena, Germany}
\affiliation{Abbe Centre of Photonics, 07745 Jena, Germany}

\author{Bernhard Urbaszek}
\affiliation{Institute of Condensed Matter Physics, Technische Universität Darmstadt, 64289 Darmstadt, Germany}

\author{Ermin Malic}
\affiliation{Department of Physics, Philipps-Universit\"at Marburg, Renthof 7, D-35032 Marburg, Germany}

\begin{abstract}
    Lateral heterostructures built of monolayers of transition metal dichalcogenides (TMDs) are characterized by a thin 1D interface exhibiting a large energy offset. Recently, the formation of spatially separated charge-transfer (CT) excitons at the interface has been demonstrated. The technologically important exciton propagation across the interface and the impact of CT excitons has remained in the dark so far. In this work, we microscopically investigate the spatiotemporal exciton dynamics in the exemplary hBN-encapsulated WSe$_2$-MoSe$_2$ lateral heterostructure and reveal a highly interesting  interplay of energy offset-driven unidirectional exciton drift across the interface and efficient capture into energetically lower CT excitons at the interface. This interplay triggers a counterintuitive thermal control of exciton transport with  a less efficient propagation at lower temperatures - opposite to the behavior in conventional semiconductors. We predict clear signatures of this intriguing exciton propagation both in far- and near-field photoluminescence experiments. Our results present an important step toward a microscopic understanding of the technologically relevant unidirectional exciton transport in lateral heterostructures.
\end{abstract}

\maketitle

\begin{figure}[t!]
    \centering
	\includegraphics[width=\columnwidth]{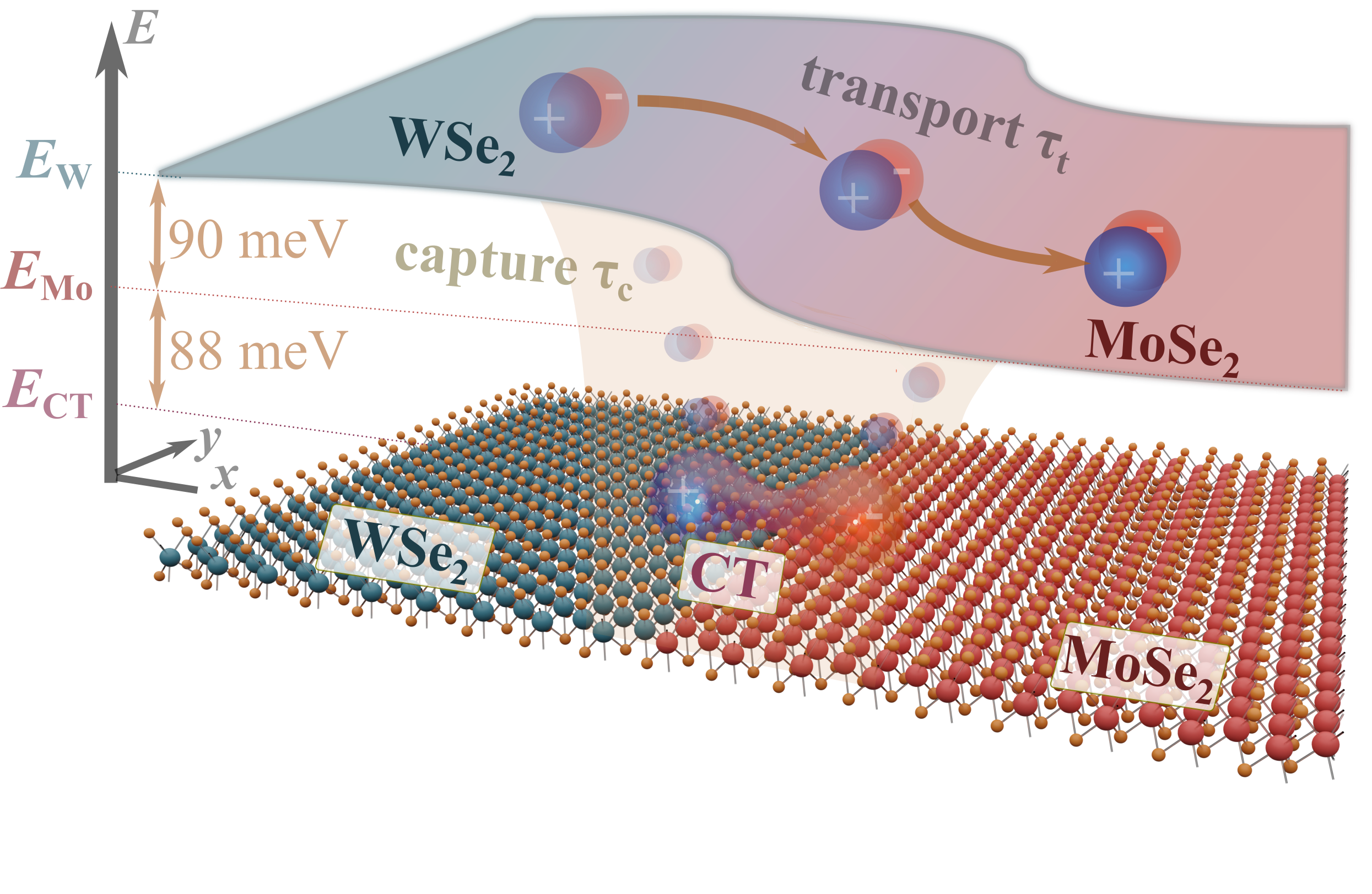}
	\caption{Sketch of a lateral TMD heterostructure (bottom) and the resulting spatially varying excitonic energy landscape (top). Exciton propagation is governed by the interplay of the  unidirectional  transport across the interface (driven by the energy offset $E_{\text{W}}-E_{\text{Mo}}$) and the capture into the energetically lowest CT excitons $X_{\text{CT}}$.
	}
	\label{fig_1}
\end{figure}

\textit{Introduction - } Monolayers of transition metal dichalcogenides (TMDs) can be grown side-by-side to form lateral heterostructures (Fig. \ref{fig_1}) \cite{Duan14,Huang14,Xie18,Sahoo18}. They  typically exhibit a band offset at the interface, facilitating the formation of  spatially separated charge-transfer (CT) excitons \cite{Lau18,Rosati23,Yuan23,Durnev24}. The latter have been experimentally demonstrated \cite{Rosati23,Yuan23}, 
 thanks to the recent developments in CVD growth techniques \cite{Najafidehaghani21,Rosati23,Beret22,Li15,Xie18,Ichinose22}, allowing narrow interface widths comparable to the exciton Bohr radius.  
 First studies on exciton and photo-carrier transport in lateral heterostructures have been performed \cite{Beret22,Yuan23,Lamsaadi2023,Bellus18,Shimasaki22,Kundu24,Zhong24,Kundu24b}. Furthermore, in near-field spectroscopy narrow exciton distributions have been realized allowing the
   observation of an unidirectional exciton drift across the interface in a WSe$_2$-MoSe$_2$ lateral heterostructure \cite{Beret22,Lamsaadi2023}.
   
   Excitons are neutral quasiparticles and their directional transport is challenging and has so far been mainly controlled in TMD monolayers or vertical TMD heterostructures, through strain profiles \cite{Rosati21e} or electric field \cite{Unuchek18}. In contrast, in lateral heterostructures, directional exciton transport occurs naturally due to the internal energy offset (Fig.  \ref{fig_1}) between the bands of the constituent materials.
    However, the role of CT excitons on the transport behaviour has been neglected so far, although they are expected to play a crucial role when excitons propagate across the interface. 
   
In this work, we  microscopically model the spatiotemporal exciton dynamics in lateral TMD heterostructures focusing, in particular, on the exemplary hBN-encapsulated WSe$_2$-MoSe$_2$ structure. We reveal an intriguing interplay between the energy offset-driven unidirectional exciton propagation across the interface and the capture into the energetically lowest CT excitons at the interface. This results in an unexpected temperature dependence with a more pronounced exciton propagation at higher temperatures - contrary to the behaviour in conventional semiconductors. 
We predict distinct signatures of this intriguing propagation behaviour in near-field spectroscopy experiments, as the crossing point in space of the PL intensity of MoSe$_2$ and WSe$_2$ exciton transitions is not at the lateral crystal junction but a distance $\Delta x $ several hundred nm away from it - in an excellent agreement with experimental data.
Overall, our microscopic and material-specific approach sheds light on exciton propagation across the interface of technologically promising lateral TMD heterostructures, in particular emphasizing the importance of CT excitons.\\

\textit{Microscopic Model - } We  first microscopically calculate the exciton energy landscape in the hBN-encapsulated WSe$_2$-MoSe$_2$ lateral heterostructure by numerically solving the Schr\"odinger equation. We consider an interface width of 2.4 nm as realized in recent CVD-grown samples \cite{Rosati23} and include the Coulomb interaction via a Keldysh-Rytova potential \cite{Keldysh79,Rytova67}. We make use of the large difference between relative and total exciton mass \cite{Kormanyos15} to separate the  Schr\"odinger equation into two coupled equations (cf. the SI) \cite{Lau18,Rosati23}. The first is a
 Wannier-like equation in the relative electron-hole position, determining if electrons and holes are bound. The solution enters the second equation, which provides an additional quantization  in the center-of-mass position \cite{Rosati23,Lau18}. This leads to the formation of CT excitons at the interface. They exhibit an energy $E_{\text{CT}}$ that is clearly lower than the energy of MoSe$_2$ and WSe$_2$ excitons $E_{\text{Mo}}$ and $E_{\text{W}}$, cf. Fig. \ref{fig_1}. The considered heterostructure has a spatial energy offset of $E_{\text{W}}-E_{\text{Mo}}$=90 meV \cite{Beret22} at the interface and  CT excitons are found to be 88 meV below $E_{\text{Mo}}$ \cite{Rosati23}.
This energy offset is expected to drive the transport of excitons from the WSe$_2$ side toward the energetically favorable MoSe$_2$ side. While excitons are crossing the interface they are likely to be  trapped into the energetically lower CT exciton states (Fig. \ref{fig_1}). 

Along the interface there is no energy offset, resulting in a drift-less diffusion  \cite{Yuan23,Voegele09}, while the offset-driven propagation across the interface can be described via one-dimensional Wigner functions $f_{\alpha}(x,q_x,t)=1/L_y \int \sum_{q_y} \sum_{\textbf{q}^\prime} \langle \hat{X}^\dagger_{\alpha,\textbf{q}+\textbf{q}^\prime/2} \hat{X}_{\alpha,\textbf{q}-\textbf{q}^\prime/2} \rangle e^{\imath \textbf{q}^\prime\cdot \textbf{r}}dy $. 
Here, $x$ and $q_x$ are the  components across the interface (which is set to be along the y direction) of the center-of-mass position $\textbf{r}=(x,y)$ and momentum $\textbf{q}=(q_x,q_y)$, while $L_{x,y}$ is the system length across/along the interface.  The index $\alpha$ = W, Mo refers to excitons at the WSe$_2$ and the MoSe$_2$ side of the heterostructure, respectively, while $\hat{X}^{(\dagger)}_{\alpha,\textbf{q}}$ are exciton annihilation (creation)  operators \cite{Rosati20, Brem18}. Using the Heisenberg equation of motion, we derive the spatiotemporal dynamics of the Wigner function yielding
\begin{equation}\label{wigner1D}
         \dot{f}_\alpha =-v_{q_x}\frac{\partial}{\partial_x} f_\alpha + \frac{f^\circ_\alpha-f_\alpha}{\tau_p} +
          \dot{f}_\alpha^{\text{d}}+ \dot{f}_\alpha^{\text{cap}}.
 \end{equation} 
 The first term describes 
 the regular exciton propagation driven by the group velocity $v_{q_x}=\hbar q_X/M$ (with the exciton mass $M$) and the gradient in the occupation. The second term describes the phonon-driven thermalization toward the local Boltzmann distribution $f^\circ_\alpha$. This is modeled with a relaxation-time approximation with the microscopically obtained phonon-driven scattering rate $\tau_p$ \cite{Selig16}. 
 As initial exciton occupation, we choose a Gaussian distribution centered at the interface or varied along the heterostructure (cf. the SI for more details).
 
 While the first two contributions are also present in regular TMD monolayers, the third and the fourth term in Eq. (\ref{wigner1D})  are specific to lateral TMD heterostructures. 
  The third term describes the drift of excitons driven by the energy offset at the interface and reads
 \begin{equation}\label{1Ddrift}
     \dot{f}_\alpha^{\text{d}} (x, q_x, t)=  \sum_{q_x^\prime} \mathcal{V}(x,q_x-q_x^\prime) f_\alpha(x,q_x^\prime, t).
 \end{equation}
 The spatial variation of exciton energies $E_\alpha(x)$ induces the superpotential $\mathcal{V}(x,q_x)=-\frac{\imath}{\hbar L_x}\int_{-x_V}^{x_V} dx^\prime \left( E_\alpha ( x+\frac{x^\prime}{2} ) - E_\alpha( x-\frac{x^\prime}{2}) \right) e^{-\imath q_xx^\prime}$, which  corresponds to the regular semiclassical drift $1/\hbar\partial_x E_\alpha(x) \partial_{q_x} f_\alpha$ for smooth spatial variations of $E_\alpha$ (cf. the SI) \cite{Hess96}. 
Finally, the capture of optically excited MoSe$_2$/WSe$_2$ excitons  into the energetically lower lying CT exciton states during the propagation across the interface is described by the forth term in Eq. \eqref{wigner1D} reading
\begin{equation}\label{cap}
        \dot{f}_\alpha^{\text{cap}} =  \frac{\Delta n_\alpha f^\circ_\alpha-f_\alpha}{\tau_c}. 
\end{equation}
Here, we have introduced
$\Delta n_\alpha=\frac{n^{\circ}_\alpha(x,t)}{n_\alpha(x,t)}$ as the ratio between the spatiotemporal exciton density $n_{\alpha}(x,t)=1/L_x\sum_{q_x}f_\alpha(x,q_x,t)$ and its thermalized analogous $n^{\circ}_\alpha(x,t)$ after introduction of CT excitons.
The capture process  is driven by the emission of optical phonons \cite{Glanemann05,Reiter07} with microscopically calculated scattering times $\tau_c$ \cite{Selig16}. 
The consequence is a decrease of  WSe$_2$ and MoSe$_2$ exciton density $n_\alpha$  and the build-up of a CT exciton  density $n_{\text{CT}}(x,t)$ at the interface.
Note that the capture is strongly temperature dependent via  the local thermalized exciton density $n^{\circ}_\alpha$. At higher temperatures, the efficiency of the capture process is reduced, as excitons can escape by absorption of phonons \cite{Reiter07}.  

The knowledge of  exciton densities allows us to microscopically model the spatially and temporally resolved photoluminescence (PL), which is determined by a product of the exciton density and the oscillator strength of the involved exciton species. The PL intensity is described by an Elliot formula \cite{Koch06,Brem20}, which we have generalized to include CT excitons \cite{Rosati23}
\begin{equation}
I(E,x,t)=\sum_\alpha  \frac{f^{\circ}_{\alpha}(x,\textbf{q}=0,t) \,\tilde{\gamma}_\alpha \left(\tilde{\gamma}_\alpha+\Gamma\right)}{(E-E_\alpha)^2+(\tilde{\gamma}_\alpha+\Gamma)^2}
\end{equation}
with the exciton index $\alpha=\text{W, Mo and CT}$  and $f^{\circ}_{\alpha}$ as the corresponding equilibrium exciton distribution. Furthermore, the PL is influenced by the temperature-dependent exciton-phonon scattering rate $\Gamma=\hbar/2\tau_p$   and  the radiative decay rate $\tilde{\gamma}=\gamma_\alpha\vert \psi_{\alpha, \textbf{q}=0}\vert^2$. Here $\gamma_\alpha$ is the oscillator strength, which is proportional to the probability of finding electrons and holes in the same position. This is about 35 smaller for spatially separated CT excitons compared to regular MoSe$_2$ and WSe$_2$ excitons \cite{Rosati23}, with the dipole furthermore having influence on the binding energy in analogy to the case of interlayer excitons \cite{Rosati23,Latini17}. The momentum conservation implies that only the $\textbf{q}=0$ component of the exciton wavefunction $\psi_{\alpha, \textbf{q}}$ allows radiative recombination.\\

 \textit{Spatiotemporal exciton dynamics - } In TMD monolayers, localized exciton densities propagate in space preserving their shape, which at low excitations only becomes spatially broader due to diffusion \cite{Kulig18}. Exciton drift can be obtained by strain engineering  \cite{Rosati21e,Harats20,Cordovilla18,Datta22,Lee22,Nysten24,Kumar24} or by applying electric fields in the case of vertical TMD heterostructures, where dipolar interlayer excitons are involved \cite{Unuchek18,Tagarelli23,Hagel23}.
In lateral TMD heterostructures, the transport behavior is drastically different with qualitative changes in the shape of the propagating exciton densities as a result of an intriguing interplay of an energy  offset-driven unidirectional exciton transport across the interface and capture into energetically lower CT excitons at the interface. 

\begin{figure}[t!]
    \centering
	\includegraphics[width=\columnwidth]{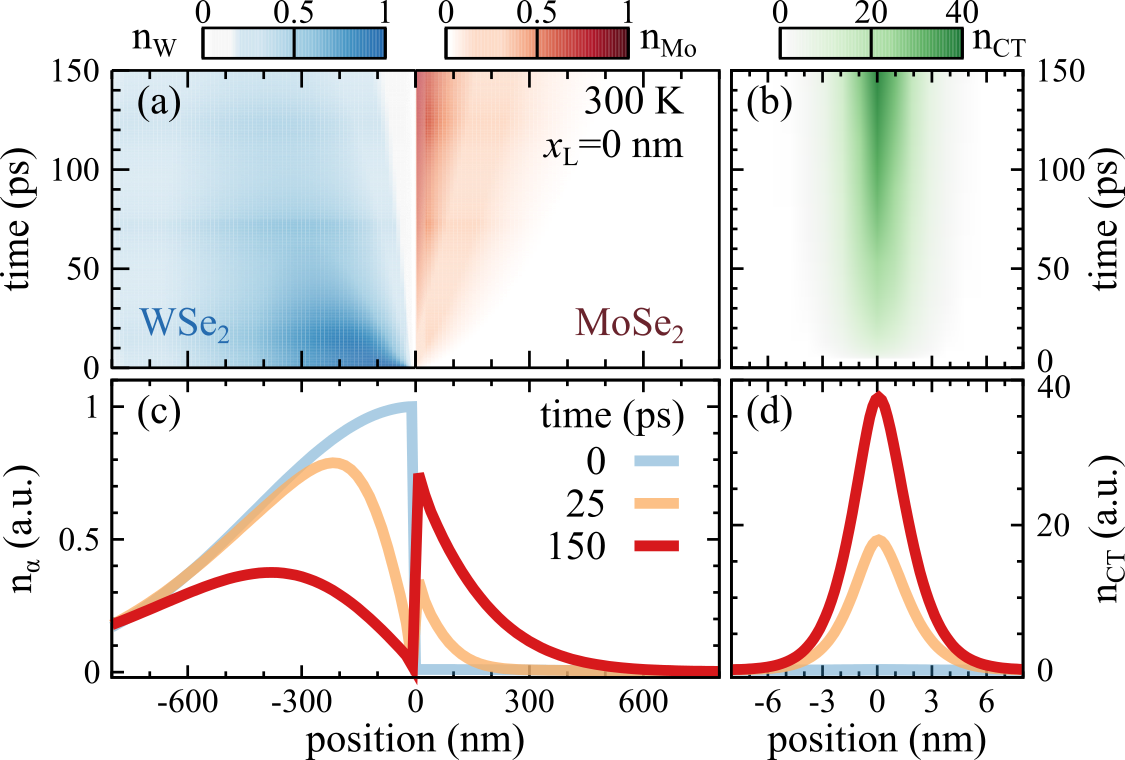}
	\caption{Space-resolved densities of (a) WSe$_2$ and MoSe$_2$ excitons as well as of  (b) CT excitons  at 300 K after an optical excitation at the interface resonant with $E_{\text{W}}$. (c)-(d) 2D cuts of exciton densities at given times. The large CT exciton occupation and the exciton accumulation at the MoSe$_2$ side of the interface can be traced back to the interplay between unidirectional exciton drift across the interface and capture into CT excitons. }
	\label{fig_2}
\end{figure}

We investigate the exciton transport in hBN-encapsulated WSe$_2$-MoSe$_2$ lateral heterostructures by numerically evaluating Eq. (1), which explicitly takes into account the interplay of drift and capture processes. Figure \ref{fig_2} shows the spatially and temporally resolved densities of (a, c) MoSe$_2$ and WSe$_2$ excitons as well as of (b, d) CT excitons after an optical excitation at the interface (laser position at $x_{\text{L}}=0$ nm) with a typical diffraction-limited spot (FWHM of 1 $\mu$m). The excitation is set to be resonant to the WSe$_2$ exciton and creates an initially broad spatial exciton density $n_{\text{W}}$ at the WSe$_2$ side of the lateral heterostructure, with a small off-resonant occupation of MoSe$_2$ excitons (about 100 times smaller than for WSe$_2$) \cite{Shree18,Chow17}, cf. the  blue line in  Fig. \ref{fig_2}(c). 
This changes rapidly, as the energy offset $E_{\text{W}}-E_{\text{Mo}}=90$ meV at the interface  gives rise to a  drift of excitons toward the energetically more favorable   MoSe$_2$ side (Fig. \ref{fig_1}). 
 This unidirectional transport leads to an accumulation of exciton density at the MoSe$_2$ side of the interface, cf. the  red line in  Fig. \ref{fig_2}(c).  The exciton accumulation becomes visible after few tens of picoseconds and it extends over hundreds of nanometers into the MoSe$_2$ side. Such broad spatial extensions are already observable in experimental setups with a spatial resolution of a few hundreds of nanometers \cite{Beret22,Yuan23,Kulig18,Cordovilla19,Rosati21e,Vazquez24}.

During the unidirectional  propagation across the interface, excitons can become trapped into the energetically lower-lying CT excitons. This results in a local depletion of the the exciton density $n_{\text{W}}$ at the interface, cf. the orange and red line in  Fig. \ref{fig_2}(c).  At the same time the density of  CT excitons considerably increases (Figs.  \ref{fig_2}(b,d)). Its maximum  becomes almost two orders of magnitude larger than the excited density within the first 150 ps. 
The efficient trapping into CT excitons has important fundamental and technological applications. Since CT excitons are dipolar and repel each other, the quick increase of their density can explain the recently observed nonlinear, dipole-driven CT-exciton diffusion along the interface \cite{Yuan23}. Since CT excitons have a small binding energy of few tens of meV \cite{Rosati23}, the trapping of WSe$_2$ or MoSe$_2$ excitons could lead to their efficient dissociation, which is favorable for many optoelectronic devices. 

In a nutshell, the drift-induced unidirectional propagation results in an exciton accumulation at the low-energy side of the interface and the efficient capture processes give rise to large CT exciton densities. The formation time and the height of exciton accumulations are determined by the interplay of exciton capture and drift and their competition can be thermally controlled.\\

\begin{figure}[t!]
    \centering
	\includegraphics[width=\columnwidth]{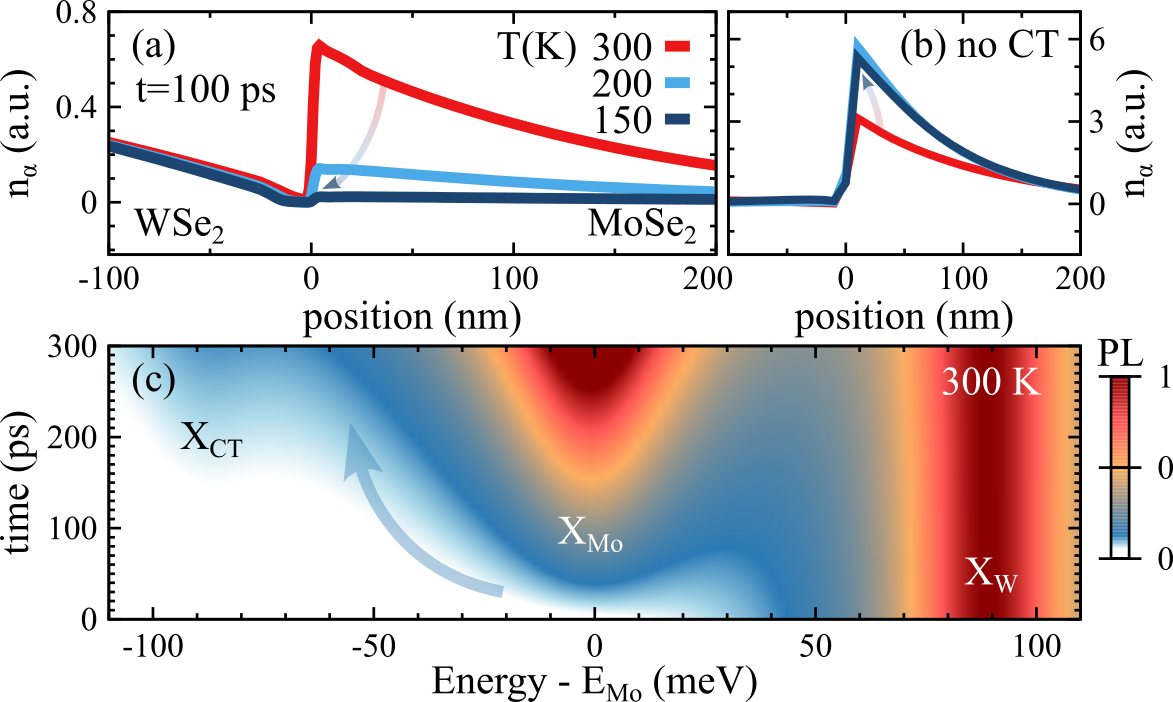}
	\caption{(a) Temperature-dependent accumulation of exciton density  after 100 ps. The decrease as a function of  temperature is opposite to the behaviour in conventional semiconductors, which we obtain when (b) CT excitons are switched off.  (c) Space-integrated and time-resolved PL spectrum (normalized to the intensity of X$_{\text{W}}$), with a delayed formation of the X$_{\text{Mo}}$ resonance. The latter is due to the energy offset-driven propagation across the interface.
	}
	\label{fig_3}
\end{figure}

\textit{Temperature control of exciton propagation - }
Carrier drift becomes generally faster at lower temperatures as here the  efficiency of scattering with phonons is decreased. 
Interestingly, we find the opposite behavior for exciton propagation in lateral TMD heterostructures. Figure \ref{fig_3}(a) illustrates the temperature-dependent spatiotemporal dynamics of MoSe$_2$ and WSe$_2$ excitons. The height of the drift-induced exciton accumulation at the MoSe$_2$ side becomes surprisingly  smaller with decreasing temperature $T$, i.e. the unidirectional transport becomes weaker at lower $T$.
This behavior can be traced back to the efficient trapping process into CT excitons at the interface. It suppresses the drift by capturing the propagating WSe$_2$ excitons before they reach the MoSe$_2$ side. To show this, we consider the case without CT excitons (corresponding to lateral heterostructures with larger interface widths \cite{Rosati23}). Here, we find the expected temperature trend, i.e. a faster propagation toward the low-energy side at smaller temperatures, cf. Fig. \ref{fig_3}(b). Note that the accumulation is overall one order of magnitude higher without CT excitons. The larger impact of the capture at smaller temperatures is due to a suppression of the counteracting escape processes driven by phonon absorption. 

The exciton accumulation at the MoSe$_2$ side after a  resonant excitation of WSe$_2$ excitons is a hallmark for the unidirectional exciton transport. This can be revealed in space-integrated and time-resolved PL, cf. Fig. \ref{fig_3}(c). The spectrum initially shows only one peak X$_{\text{W}}$ stemming from optically excited WSe$_2$ excitons centered at $E_{\text{W}}$. The energy offset-induced exciton drift results in the formation of an additional resonance  at  $E_{\text{Mo}}$ after a certain time delay.  While the linewidth of this peak is fixed by radiative-recombination and phonon-mediated scattering  \cite{Selig16}, its height relative to X$_{\text{W}}$ increases in time due to the unidirectional exciton propagation.
At room temperature, X$_{\text{Mo}}$ becomes as intense as X$_{\text{W}}$ after approximately 200 ps. The situation is drastically different at smaller temperatures, where the capture-induced suppression of exciton propagation results in a negligible intensity of the drift-driven X$_{\text{Mo}}$ resonance (cf. the SI). Note that an excitation resonant to $E_{\text{Mo}}$  does not result in the formation of a X$_{\text{W}}$ peak, as excitons would need to increase their energy by about 90 meV.
This indicates the possibility of an optical control of the unidirectional exciton drift that is potentially interesting for excitonic diodes \cite{Butov17,Beret22}. 

Surprisingly, we also observe the appearance of a small CT exciton peak X$_{\text{CT}}$ after few hundreds of picoseconds even at room temperature (Fig. \ref{fig_3}(c)). The peak becomes more intense at reduced temperatures in particular compared to the X$_{\text{Mo}}$ resonance. The formation time of the CT exciton resonance is, however,  close to typical PL decay times of 150-200 ps in  WSe$_2$-MoSe$_2$ lateral heterostructures \cite{Beret22,Najafidehaghani21}. As a consequence, such a small CT exciton peak present in time-resolved PL spectra might be difficult to observe in time-integrated, energy-resolved far-field PL. Thus, we investigate now optical excitation in the nanometer range as can be realized in near-field spectroscopy.\\

\begin{figure}[t!]
    \centering
	\includegraphics[width=\columnwidth]{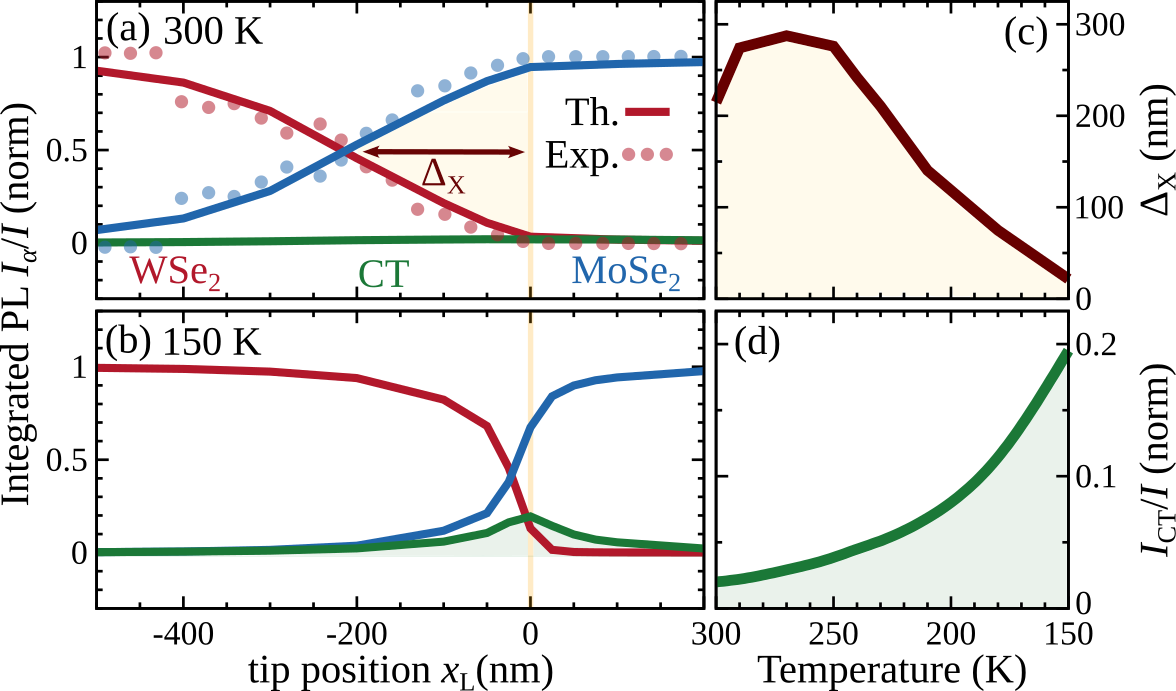}
	\caption{Intensity of the X$_{\text{Mo}}$, X$_{\text{W}}$ and X$_{\text{CT}}$ resonances in PL spectra at (a) 300 K and  (b) 150 K after a near-field excitation at different laser positions (normalized to the total intensity $I$). The vertical yellow line indicates the position of the interface. The dots show the corresponding experimental measurement taken from Ref. \cite{Beret22}. The interplay of the energy offset-driven unidirectional exciton propagation and capture into CT excitons breaks the symmetry resulting in a spatial offset $\Delta_{\text{X}}$ from the interface, at which the emission from WSe$_2$ and MoSe$_2$ excitons is equal - in excellent agreement between theory and experiment. (c)-(d) Temperature-dependent offset $\Delta_{\text{X}}$ and CT exciton emission $I_{\text{CT}}/I$.
	}
	\label{fig_4}
\end{figure}

\textit{Near-field photoluminescence - } 
In near-field  scanning microscopy, we focus on excitations spots with FWHM of 50 nm and varying position in space. In this way, we create asymmetric MoSe$_2$ and WSe$_2$ exciton occupations as realized in recent experiments  \cite{Beret22,Albagami22,Lamsaadi2023}.
Figure \ref{fig_4}(a) shows the relative space- and time-integrated PL intensity as a function of the laser position  across the interface.
We consider an exciton decay time of 175 ps as extracted from measurements on the same lateral heterostructure \cite{Beret22}.
When exciting at the MoSe$_2$ side, the PL is, as expected, clearly dominated by I$_{\text{Mo}}$ and the I$_{\text{W}}$ signal is negligible, cf. the blue and red line in Fig. \ref{fig_4}(a).

 Naively, we would expect excitations on the WSe$_2$ side to induce PL dominated by the X$_{\text{W}}$ resonance. The predicted PL is, however, drastically different due to energy offset-driven unidirectional transport to the MoSe$_2$ side and the capture processes into CT excitons at the interface. 
When the excitation spot is far away from the interface ($>$500 nm), the PL behaves as expected and I$_{\text{W}}$ dominates. However, for excitation closer to the interface ($<$200 nm),  I$_{\text{Mo}}$ becomes more pronounced than I$_{\text{W}}$. Note that this occurs for excitation spots with  a distance 4 times larger than the excitation confinement of 50 nm. 
The reason for this behavior is the efficient exciton drift, which drives the excited WSe$_2$ excitons toward the energetically more favorable MoSe$_2$ side. The drift efficiency can be quantified by the spatial offset $\Delta_{\text{X}}$ from the interface with an equal PL intensity from MoSe$_2$ and WSe$_2$ excitons. Without a drift, we find $\Delta_{\text{X}}$ to be 0, as expected. At room temperature, we predict a value of $\Delta_{\text{X}}\approx 215 $nm. This is in excellent agreement with recent near-field experiments \cite{Beret22,Lamsaadi2023}, cf. the dots in Fig. \ref{fig_4}(a).

Given the thermal control of exciton transport discussed above, we investigate the near-field PL at the lower temperature of 150 K (Fig. \ref{fig_4}(b)). We observe that  the spatial offset becomes smaller with $\Delta_{\text{X}}\approx 20$nm reflecting the capture-induced suppression of the exciton drift at decreasing temperatures (cf. also Fig. \ref{fig_3}(a)). This is confirmed by a temperature study in Fig. \ref{fig_4}(c), where we interestingly also find a slight increase of $\Delta_{\text{X}}$ for temperatures from 300 to 280 K. This is induced by the competition between the intrinsic energy offset-driven exciton drift and exciton capture (both more efficient at smaller temperatures). The efficient capture is further demonstrated in the increasing PL intensity of CT excitons at lower temperatures,  cf. Fig. \ref{fig_4}(d). Here, the near-field excitation allows to visualize  the CT resonance even in time-integrated PL. \\

\textit{Conclusions - } Based on a microscopic, material specific and predictive approach, we have studied exciton transport in lateral TMD heterostructures. We demonstrate a pronounced 
unidirectional exciton drift due to the energy offset across the interface resulting in an accumulation of excitons at one side of the interface. Furthermore, we predict a crucial impact of capture processes into the energetically lower-lying charge transfer excitons at the interface. Finally, we demonstrate that temperature is the key knob to control the interplay between the  unidirectional exciton drift and the capture-induced formation of charge transfer excitons. 
Based on the gained microscopic knowledge, we provide concrete recipes for detecting the intriguing exciton propagation  both in far- and near-field photoluminescence experiments. Our findings have also a potential technological importance for devices based on  unidirectional exciton transport in lateral heterostructures.\\

\begin{acknowledgments}
We  acknowledge support by the Deutsche Forschungsgemeinschaft   (DFG) via SPP 2244 and CRC 1083 (project B9). We also thank Giuseppe Meneghini for graphical support in Fig. \ref{fig_1} and for discussions together with Jamie Fitzgerald.
\end{acknowledgments}

\end{document}


\title{Supplementary Information for \\ Impact of charge transfer excitons on unidirectional exciton transport \\ in lateral TMD heterostructures}

\author{Roberto Rosati}
\email{roberto.rosati@physik.uni-marburg.de}
\affiliation{Department of Physics, Philipps-Universit\"at Marburg, Renthof 7, D-35032 Marburg, Germany}

\author{Sai Shradha}
\affiliation{Institute of Condensed Matter Physics, Technische Universität Darmstadt, 64289 Darmstadt, Germany}

\author{Julian Picker}
\affiliation{Institute of Physical Chemistry, Friedrich Schiller University Jena, 07743 Jena, Germany}
\affiliation{Abbe Centre of Photonics, 07745 Jena, Germany}

\author{Andrey Turchanin}
\affiliation{Institute of Physical Chemistry, Friedrich Schiller University Jena, 07743 Jena, Germany}
\affiliation{Abbe Centre of Photonics, 07745 Jena, Germany}

\author{Bernhard Urbaszek}
\affiliation{Institute of Condensed Matter Physics, Technische Universität Darmstadt, 64289 Darmstadt, Germany}

\author{Ermin Malic}
\affiliation{Department of Physics, Philipps-Universit\"at Marburg, Renthof 7, D-35032 Marburg, Germany}

\maketitle



\section{Exciton energy landscape}\label{Sec:theory-landscape}

\begin{figure}[t!]
    \centering	\includegraphics[width=0.5\columnwidth]{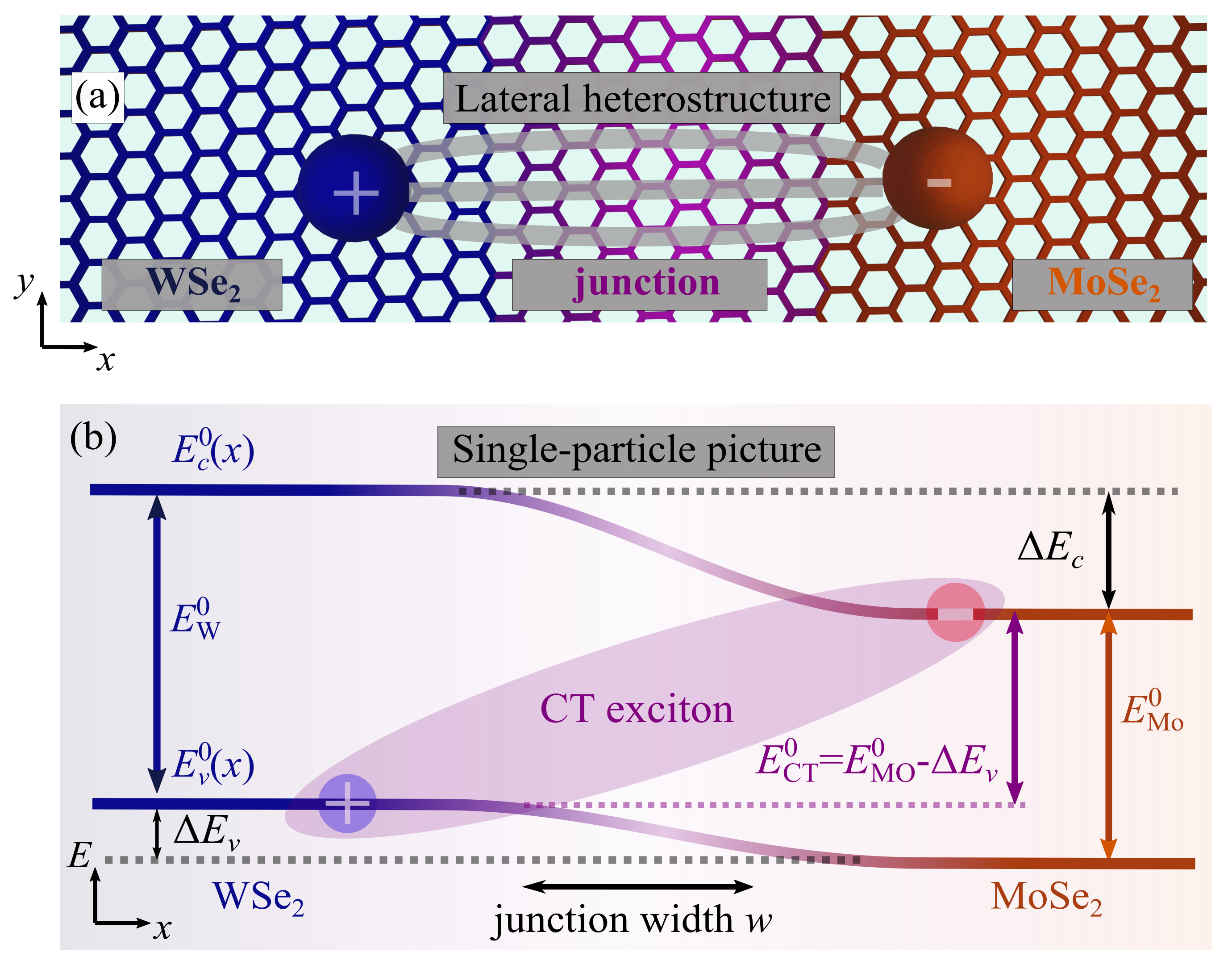}
	\caption{(a) Sketch of a lateral  WSe$_2$-MoSe$_2$ heterostructure. (b) The two materials have intrinsic bandgaps $E^0_{\text{Mo}}$ and $E^0_{\text{W}}$ and forming conduction and valence band offsets $ \Delta E_c, \Delta E_v$ around the junction. Spatially-separated  electrons and holes across the interface form charge-transfer (CT) excitons with the corresponding continuum energy E$^0_{\text{CT}}=E^0_{\text{Mo}}-\Delta E_v$.
	}
	\label{fig_1}
\end{figure}

We investigate lateral heterostructures (LHs) based on  transition metal dichalcogenide monolayers (TMD). We solve the Schr\"odinger equation including the  space-dependent energy landscape due to the presence of an interface, cf.  Fig. S\ref{fig_1}. The two different materials induce an in-plane variation of energy, which we describe via the single-particle band-edge energies $E^0_{c/v}(\mathbf{r}_{c/v})$ of conduction- and valence-band electrons, respectively, with $\mathbf{r}_{c/v}$ being their respective positions. These single-particle energies read 
$E^0_{c/v}(\mathbf{r}_{c/v})=\Delta E_{c/v} (1-\text{tanh}(4x_{c/v}/w))/2+E^0_{\text{Mo}}(1\pm 1)/2$ with $+$ ($-$) for $c\, (v)$, cf. Fig. S\ref{fig_1}(b) \cite{Lau18,Rosati23}.  Here,  $w$ denotes a finite width of the interface, reflecting the typical situation in real samples due to spontaneous alloying \cite{Duan14,Huang14}. In particular, we consider a narrow interface of $w$=2.4 nm, as recently realized experimentally \cite{Rosati23}.  While lateral heterostructures involving TMDs with different chalcogen atoms have a  lattice mismatch, in WSe$_2$-MoSe$_2$ we can assume a strain-free interface \cite{Zhang18,Xie18}.   The conduction and valence bands form an offset $ \Delta E_c, \Delta E_v$ at the interface, typically inducing a type-II alignment \cite{Kang13,Chu18,Sahoo18,Najafidehaghani21,Herbig21}. In  the specific case of WSe$_2$-MoSe$_2$ LH, the conduction band minimum is located in the MoSe$_2$ layer \cite{Guo16}, hence only the bright valley is relevant for charge transfer (CT) excitons in WSe$_2$-MoSe$_2$ \cite{Rosati23}.
We take a valence-band offset $\Delta E_v$=215 meV in agreement with the microscopic estimation in similar LHs \cite{Guo16}. In addition, the two materials exhibit also a different bandgap with $E^0_{\text{W}}-E^0_{\text{Mo}}=90$ meV \cite{Beret22} leading to $\Delta E_c= \Delta E_v + 90$ meV. The different bandgap at the two sides of the heterostruture is different than in the case of gate-induced homojunctions \cite{Pospischil14,Baugher14,Ross14}, where the band offsets are the same ($\Delta E_c= \Delta E_v$) and where bound excitons confined to few tens of nanometers can appear \cite{Thureja22,Heithoff24}. 
The type-II alignment allows for a continuum of unbound CT electron-hole pairs, with electrons and holes stemming from different sides of the interface and forming an unbound continuum with the minimum  energy $E_{\text{CT}}^0=E^0_{\text{Mo}}-\Delta E_v= E^0_{\text{W}}-\Delta E_c$  (cf. the purple oval in Fig. S\ref{fig_1}(b)).

To investigate if also bound CT excitons can form, we  include the electron-hole Coulomb interaction $V_C(\textbf{r}_r)$, which is described by 
a generalized Keldysh potential \cite{Rytova67,Keldysh79,Brem19b} for charges 
in a thin-film surrounded by a spatially-homogeneous dielectric environment  \cite{Rytova67,Keldysh79,Brem19b}. Here we have introduced the center-of-mass  and relative positions $\textbf{r}\equiv(x,y)$ and $\textbf{r}_{r}$, respectively.
In view of the difference between the relative and total exciton mass $\mu$ and $M$, respectively, the resulting Schr\"odinger equation can be separated in equations for the relative and the center-of-mass motion \cite{Lau18,Rosati23} yielding
\begin{subequations}\label{sepHam}
\begin{align}
         \left[\!\frac{-\hbar^2\nabla^2_{\textbf{r}_r}}{2\mu} \!+\! V_C(\textbf{r}_r) \!\!+\!\! V^{x}\!(\textbf{r}_r) \!\right] \!\!\phi^{x}_j\!(\textbf{r}_r)\!\!=\!\!\tilde{E}_j(x) \phi^{x}_j\!(\textbf{r}_r) \,,  \label{Wannier}  \\
        \left[-\frac{\hbar^2}{2M}\partial^2_{x} + \tilde{E}_j(x) \right]\psi_{n,j}(x)= E_{n,j}\psi_{n,j}(x) \,, \label{CM}
    \end{align}
\end{subequations}
where $V^{x}(\bm r_r)=E^0_c( \bm{r}_r, x)-E^0_v(\bm{r}_r, x)$ acts as an interface potential  given by the space-dependent band edges $E^0_{c,v}$ and the masses $\mu$ and $M$, which we include as  material-specific input parameters from first-principles calculations \cite{Kormanyos15}. 
Here, Eq. (\ref{Wannier}) is a Wannier-like equation stating if electrons and holes can bind together via a quantization in relative motion, with $j=$1s, 2s, 3s, ... labeling the bound states. In this work, we focus on the energetically lowest 1s excitons justified by the  large energy separation of excited states. The eigenenergies $\tilde{E}_j(x)$ found by solving  Eq. (\ref{Wannier}) act as a potential in Eq. (\ref{CM}), where they can induce an additional quantization in the center-of-mass position, resulting in the emergence of bound CT excitons localized at the interface. 
In the monolayer limit of no band offsets, i.e.  $ \Delta E_{c/v}=0$, 
Eq. (\ref{Wannier}) becomes the well-known Wannier equation with a space-independent potential  and $\tilde{E}_j(x)\equiv \tilde{E}_j$.
 In this limit, the  center-of-mass equation (Eq. (\ref{CM})) becomes trivial corresponding to fully delocalized plane waves $\psi_n(x)\equiv \psi_{q_x}(x)= e^{\imath q_{x} x}$ and resulting in  $E_{n,j}\equiv E_{q_x,j}=\tilde{E}_j+\hbar^2 q_{x}^2/2M$. Here, $q_x$ is the component across the interface of the center-of-mass momentum $\textbf{q}=(q_x,q_y)$. This implies that the center-of-mass motion of excitons is free and there is no quantization. For WSe$_2$-MoSe$_2$ LHs, we find CT excitons with an energy E$_{\text{CT}}\equiv E_{1,1s}$ that is approximately 88 meV lower than  E$_{\text{Mo}}\equiv \tilde{E}_{1s}(x\gg 0)$ (and 178 meV lower than  E$_{\text{W}}\equiv \tilde{E}_{1s}(x\ll 0))$. While spatially localized across the interface, CT excitons are delocalized along the junctions, where states with the momentum $q_y$ along the interface form a one-dimensional parabolic dispersion. Overall, the wave-function of CT excitons  reads $\Psi_{q_y}(\textbf{r},\textbf{r})_r=\psi_{\text{CT}}(x)\,\phi^{x}_{1s}\!(\textbf{r}) \frac{1}{\sqrt{L_y}} e^{\imath q_y y}$, where $\psi_{\text{CT}}(x)\equiv\psi_{1,1s}$ is the ground eigenstate in Eq. (\ref{CM}) and $L_y$ is the length of the sample along the interface. 
 
 \section{Spatiotemporal exciton dynamics}\label{Sec:spatioTemporal}

 To investigate the spatiotemporal exciton dynamics in a lateral heterostructure, we first introduce a two-dimensional description of the dynamics and then exploit the sample symmetry to reduce it to the simpler one-dimensional case. For this purpose, we start with the  Wigner functions $f^{\text{2D}}_\alpha(\textbf{r},\textbf{q},t)=\sum_{\textbf{q}^\prime} \langle \hat{X}^\dagger_{\alpha,\textbf{q}+\textbf{q}^\prime/2} \hat{X}_{\alpha,\textbf{q}-\textbf{q}^\prime/2} \rangle e^{\imath \textbf{q}^\prime\cdot \textbf{r}}$ 
  with $\alpha=$W, Mo denoting WSe$_2$ and MoSe$_2$ excitons, respectively.  
Here, $\hat{X}^{(\dagger)}_{\alpha,\textbf{q}}$ are the exciton annihilation (creation)  operators \cite{Rosati20, Brem18}. Using the Heisenberg equation of motion, we derive the spatiotemporal dynamics of the Wigner function yielding
\begin{equation}\label{2Ddyn}
    \begin{split}
        \dot{f}^{\text{2D}}_\alpha(\textbf{r},\textbf{q},t)=&-v_\textbf{q}\cdot \nabla f^{\text{2D}}_\alpha(\textbf{r},\textbf{q},t)+\frac{f^{\text{2D}, \circ}_\alpha(\textbf{r},\textbf{q},t)-f^{\text{2D}}_\alpha(\textbf{r},\textbf{q},t)}{\tau_p} + \dot{f}^{\text{2D, d}}_\alpha(\textbf{r},\textbf{q},t)+\dot{f}^{2D,\text{cap}}_\alpha(\textbf{r},\textbf{q},t).
    \end{split}
\end{equation}
 The first term describes 
 the regular exciton propagation driven by the group velocity $v_{\textbf{q}}=\hbar \textbf{q}/M$  and the gradient in the occupation. The second term describes the phonon-driven thermalization toward the local Boltzmann distribution 
 \begin{equation}\label{therm2D}
  f^{\text{2D}, \circ}_{\alpha}(\textbf{r},\textbf{q},t)=A\bar{f}^{\text{2D}, \circ}_{\alpha}(\textbf{q})\, n_{\alpha}(\textbf{r},t) \quad ,   
 \end{equation}
 with $\bar{f}^{\text{2D}, \circ}_{\alpha}(\textbf{q})=\frac{2\pi\hbar^2}{Mk_BT}\exp{\left(-\frac{\hbar^2|\textbf{q}|^2}{2Mk_BT}\right)}/A$ being the normalized Boltzmann distribution, where $A=L_xL_y$ is the area of the sample.
 Here, $n_{\alpha}(\textbf{r},t)=\frac{1}{A} \sum_{\textbf{q}}f^{\text{2D}}_{\alpha}(\textbf{r},\textbf{q},t)$ 
 is the corresponding spatial exciton density. 
 The phonon-driven thermalization is typically described by the Boltzmann collision term,  which can be simplified in the relaxation-time approximation in the limit of $\sum_{\textbf{q}^\prime} \Gamma_{\alpha,\textbf{q};\alpha,\textbf{q}^\prime}(f^{\text{2D}, \circ}_{\alpha}(\textbf{r},\textbf{q}^\prime,t)-f^{\text{2D}}_{\alpha}(\textbf{r},\textbf{q}^\prime,t))\approx 0$, 
 where $\Gamma_{\alpha,\textbf{q};\alpha,\textbf{q}^\prime}$ are the scattering coefficients from exciton $\vert \alpha,\textbf{q}^\prime \rangle$ into $\vert \alpha,\textbf{q} \rangle$ \cite{Hess96}. Such scattering coefficients provide the state-dependent scattering times $\tau_{\alpha,\textbf{q}}^{-1}=\sum_{\textbf{q}^\prime} \Gamma_{\alpha,\textbf{q}^\prime;\alpha,\textbf{q}}$ \cite{Hess96}. Here, we assume a state-independent $\tau_{\alpha,\textbf{q}}=\tau_p$ with the microscopically obtained phonon-driven scattering rate $\tau_p$ provided in Ref. \cite{Selig16}. 

  While the first two terms in Eq. (\ref{2Ddyn}) are present also in regular TMD monolayers, the third and the fourth term are specific to lateral TMD heterostructures. 
  The third term describes the drift of excitons driven by the energy offset at the interface and reads
\begin{equation}\label{1DdriftVcal}
    \begin{split}
        \dot{f}_\alpha^{\text{2D,d}} (\textbf{r}, \textbf{q}, t)=-\hbar^2 \int d\textbf{q}^\prime \mathcal{V}_{\text{2D}}(\textbf{r},\textbf{q}-\textbf{q}^\prime) f^{\text{2D}}_\alpha(\textbf{r},\textbf{q}^\prime,t) = -\hbar \int d\textbf{q}^\prime \delta(q_y-q^\prime_y)\mathcal{V}_{\text{1D}}(x,q-q^\prime) f^{\text{2D}}_\alpha(\textbf{r},\textbf{q}^\prime,t). 
    \end{split}
\end{equation}    
Here, $\mathcal{V}$ is induced by the excitonic energy offset and obtained by introducing the interface potential $E_\alpha$ (cf. Sec. \ref{Sec:theory-landscape}) \cite{Rosati14b}. For the specific case, it reads
  \begin{subequations}\label{Vcal}
     \begin{align}
        \mathcal{V}_{\text{2D}}(\textbf{r},\textbf{q}^{\prime\prime})=\frac{i}{(2\pi)^2\hbar^3}\int d\textbf{r}^\prime \left[ E_\alpha\left( \textbf{r}+\frac{\textbf{r}^\prime}{2} \right) - E_\alpha\left( \textbf{r}-\frac{\textbf{r}^\prime}{2} \right) \right] e^{-\imath \textbf{q}^{\prime\prime}\cdot \textbf{r}^\prime}  = \frac{\delta(q_y^{\prime\prime})}{\hbar} \mathcal{V}(x,q_x^{\prime\prime}) \,, \label{Vcal-gen-1D} \\[12pt]
        \text{with } \mathcal{V}_{\text{1D}}(x,q_x^{\prime\prime})=\frac{i}{2\pi\hbar^2}\int_{-x_V}^{x_V} dx^\prime \left[ E_\alpha\left( x+\frac{x^\prime}{2} \right) - E_\alpha\left( x-\frac{x^\prime}{2} \right) \right] e^{-\imath q_x^{\prime\prime}x^\prime} \quad , \label{Vcal1D}
     \end{align}
 \end{subequations}
 where we used the property of the LH interface potential to be invariant along the interface  ($E_\alpha\left( \textbf{r} \right)\equiv E_\alpha\left( x \right)$) and introduced a reasonable cutoff of $x_V=$25 nm.
 Taylor expanding $E_\alpha\left( x\pm\frac{x^\prime}{2}\right)$ around $x$ and after using $x^{\prime n}e^{-\imath q^{\prime\prime}x^\prime}=(-1/\imath)^n \partial^n/\partial q^{\prime\prime n}e^{-\imath q^{\prime\prime}x^\prime}$ together with a partial integration, one can show that \cite{Hess96}
 \begin{equation}\label{expansion}
 \dot{f}_\alpha^{\text{2D, d}} (\textbf{r}, \textbf{q}, t)\approx \sum_n c_n \nabla^n_\textbf{r}E_\alpha(\textbf{r}) \cdot \nabla^n_\textbf{q} f^{\text{2D}}_\alpha(\textbf{r}, \textbf{q}, t) 
 \end{equation}
 with complex constants $c_n$. For smooth-enough energies $E_\alpha(\textbf{r})$, the first order of the expansion is enough, leading to the well-known semiclassical drift $\dot{f}_\alpha^{\text{2D,d}}= \nabla_\textbf{r}E_\alpha(\textbf{r}) \cdot \nabla_\textbf{q}f_\alpha^{\text{2D,d}}/\hbar$. However, for realistic interface energies $E_\alpha(x)$ varying within a width $w=2.4$ nm, all terms in Eq. (\ref{expansion}) should be included.
 
  Finally, the last term  in Eq. (\ref{2Ddyn}) describes the trapping of MoSe$_2$ and WSe$_2$ excitons into the bound CT states. Similarly to the case of carrier-capture \cite{Glanemann05,Reiter07,Rosati17}, this is expected to be driven by scattering with phonons and be local, i.e. taking place only when MoSe$_2$/WSe$_2$ excitons are located at the interface, where CT excitons are localized. 
In analogy to the intraband phonon-driven dynamics (second term in Eq. (\ref{2Ddyn})),
this is expressed in the relaxation time approximation as
\begin{equation}\label{cap}
        \dot{f}_\alpha^{\text{2D,cap}}  = \frac{\Delta n_\alpha f^{\text{2D}, \circ}_\alpha-f^{\text{2D}}_\alpha}{\tau_c}.
\end{equation}
Here, we have introduced $\Delta n_\alpha=\frac{n^{\circ}_\alpha(\textbf{r},t)}{n_\alpha(\textbf{r},t)}$ as the ratio between the spatiotemporal exciton density $n_{\alpha}(\textbf{r},t)$ and its thermalized density $n^{\circ}_\alpha(\textbf{r},t)$. This is obtained after the introduction of CT excitons in the  local thermalized spatial density of $\alpha=$Mo, W excitons as $n^{\circ}_\alpha(\textbf{r},t)=d_\alpha(x)\left(n_\alpha(\textbf{r},t)+n_{\text{CT}}(\textbf{r},t)\right)$ with 
 \begin{equation}\label{r0}
     d_\alpha(x)=\left(1+e^{-\frac{E_{\text{CT}}-E_\alpha(x)}{k_B T}}\left\vert \psi_{\text{CT}}(x) \right\vert^2\sqrt{\frac{2\pi\hbar^2}{Mk_BT}} \sum_{q_y} e^{-\frac{\hbar^2q_y^2}{2Mk_BT}}\right)^{-1}.
 \end{equation}
Since the capture is dominated by the emission of optical phonons, we assume $\tau_c=\tau_{\text{abs},c}e^{-\Omega/{k_BT}}/\left(1+e^{-\Omega/{k_BT}}\right)$, where $\tau_{\text{abs},c}$ is the temperature-dependent scattering time of MoSe$_2$ due to absorption of intravalley optical phonons with an energy of $\Omega=30$ meV, which has been microscopically evaluated in Ref. \cite{Selig16}. 
 Equation (\ref{cap}) implies a decrease of $n_\alpha(\textbf{r},t)$ as a consequence of trapping only when $d_{\alpha}$ differs from one. In agreement with the locality of carrier capture \cite{Glanemann05,Reiter07,Rosati17}, this takes place only where the wavefunction $\psi_{\text{CT}}(x)$ of the bound CT exciton state is finite, hence close to the interface. Furthermore, the trapping is suppressed once the ratio between $n_\alpha$ and $n_{{\text{CT}}}$ approaches the equilibrium value ($n_\alpha\approx n^\circ_\alpha$). This takes place when the CT exciton occupation is high enough to have a compensation between escape and trapping. As shown in the main manuscript, this compensation takes place on very different timescales depending on the temperature of the system. 

Along the interface there is no energy offset, resulting in a drift-less diffusion  \cite{Yuan23,Voegele09}, while we expect a propagation across the interface driven by the energy offset. This can be described by a one-dimensional Wigner function $f_\alpha(x,q_x,t)$ obtained by integrating over space- and momentum  along the interface yielding 
\begin{equation}\label{wigner1D}
    f_{\alpha}(x,q,t)= \frac{1}{L_y}\sum_{q_y} \int dy f^{\text{2D}}_{\alpha}(\textbf{r}_{CM},\textbf{q}_{CM},t)
\end{equation}
and for its temporal evolution
\begin{subequations}\label{1Ddyn}
     \begin{align}
         \dot{f}_{\alpha}(x,q_x,t)= \frac{1}{L_y}\sum_{q_y}\int dy \dot{f}^{\text{2D}}_{\alpha}(\textbf{r},\textbf{q},t) =-v_{q_x}\frac{\partial}{\partial_x} f_\alpha + \frac{f^\circ_\alpha-f_\alpha}{\tau_p} +
          \dot{f}_\alpha^{\text{d}}+ \dot{f}_\alpha^{\text{cap}} \\
     \text{with} \quad
          \dot{f}_\alpha^{\text{d}} (x, q_x, t)=  \sum_{q_x^\prime} \mathcal{V}(x,q_x-q_x^\prime) f_\alpha(x,q_x^\prime, t) \quad \text{and}\quad
          \dot{f}_\alpha^{\text{cap}} =  \frac{\Delta n_\alpha f^\circ_\alpha-f_\alpha}{\tau_c} \quad . \label{1Dtrap}
     \end{align}
 \end{subequations}
To obtain Eq. (\ref{1Ddyn}) we have assumed an identical distribution at the edges, i.e. $f^{\text{2D}}_{\alpha}\left( (x,L_y/2),\textbf{q},t \right)\approx f^{\text{2D}}_{\alpha}\left( (x,-L_y/2),\textbf{q},t \right)$ and introduced $f^\circ_{\alpha}(x,q,t)= \frac{1}{L_y}\sum_{q_y} \int dy f^{\text{2D}\circ}_{\alpha}(\textbf{r}_{CM},\textbf{q}_{CM},t)$ and $\mathcal{V}(x,q_x)=-2\pi\hbar/L_x \mathcal{V}_{\text{1D}}(x,q_x)$.  The trapping of MoSe$_2$ and WSe$_2$ excitons  forms CT excitons, whose spatial density $n_{\text{CT}}(x,t)$ has a profile across the interface determined by the wavefunction $\psi_{\text{CT}}(x)$, hence $n_{\text{CT}}(x,t)=\left\vert \psi_{\text{CT}}(x) \right\vert^2 N_{\text{CT}}(t)$, where $N_{\text{CT}}(t)$ is the time-dependent total population of CT excitons. As a consequence, Eq. (\ref{1Ddyn}) is coupled with the dynamics of the CT exciton density by
 \begin{equation}\label{dn-vsdN}
      \dot{n}_{\text{CT}}(x, t)= \left\vert \psi_{\text{CT}}(x) \right\vert^2 \dot{N}_{\text{CT}}(t),
  \end{equation}
 with the temporal evolution of $N_{\text{CT}}(t)$ following from the particle conservation as 
 \begin{equation}\label{dN}
      \dot{N}_{\text{CT}}(t)=-\frac{1}{L_x}\sum_{q_x}\sum_{\alpha=\text{Mo,W}}\int dx \dot{f}_\alpha^{\text{cap}}(x,q_x,t).
  \end{equation}

\section{Photoluminescence}\label{Sec:theory-PL}
The optical excitation of TMD-based heterostructure leads to a coherent-to-incoherent population transfer \cite{Selig18,Brem18} followed by an energy-relaxation \cite{Brem18,Wallauer21,Schmitt22} potentially relevant also for transient transport phenomena \cite{Rosati20,Rosati21c}. However, such transient effects last only few hundreds of femtoseconds at high temperatures \cite{Rosati20}, a timescale much shorter than the one investigated here. As a consequence, we start from an initial exciton distribution in the form of
 \begin{equation}\label{initial}
       f^{\circ}_{\alpha}(x,q_x,t=0)=L_x\bar{f}^{\circ}_{\alpha}(q{_x})e^{-\frac{(x-x_L)^2}{2\Delta_0^2}} g_0(x) \quad ,   
 \end{equation}
 where $\bar{f}^{\circ}_{\alpha}(q_X)=1/L_x\frac{\sqrt{2\pi}\hbar}{\sqrt{Mk_BT}}\exp{\left[-\frac{\hbar^2|\textbf{q}|^2}{2Mk_BT}\right]}$ is the normalized Boltzmann distribution, while $x_L$ and $\Delta_0$ provide the position and the size of the laser spot, respectively. Finally, $g_0$ provides  spatial inhomogeneities induced by the energy selection rule: in the case of an excitation resonant to WSe$_2$ we set
     $g_0(x)=\Theta(-x)+c_L\Theta(x) $,  with $c_L=0.01$ leading to a much larger excitation at the WSe$_2$ side ($x<0$). In view of the 90 meV separation between the laser energy and $E_{\text{Mo}}$, the MoSe$_2$ excitons are  formed less efficiently via phonon-driven non-resonant processes \cite{Shree18,Chow17}. In contrast, for the case of off-resonant excitations we expect only minor differences in the formation of WSe$_2$ and MoSe$_2$ excitons, leading to $g_0=1$.
     
The solution of Eq. (\ref{sepHam}) provides  exciton wavefunctions, which allows to determine the oscillator strength $\gamma_{\alpha}$. The latter is proportional to the probability ${|\phi(\textbf{r}=0)|^2=\int dR_x |\psi(R_x)|^2 |\phi^{R_x}(\textbf{r}=0)|^2}$ of finding electrons and holes  in the same position \cite{Rosati23,Lau18}. In the specific case of the hBN-encapsulated WSe$_2$-MoSe$_2$ lateral heterostructure, the spatial separation of CT excitons leads to an oscillator strength that is about 35 times smaller than for MoSe$_2$ or WSe$_2$ excitons \cite{Rosati23}. The spatial dipole furthermore leads to smaller binding energies is analogy to the case of interlayer excitons in vertical heterostructures \cite{Rosati23,Latini17}. The energy-, space- and time-resolved photoluminescence $I(E,x,t)$ (PL) is determined by a product of the exciton density and the oscillator strength of the involved exciton species resulting in an Elliot formula \cite{Koch06,Brem20}, which we have generalized to include CT excitons  \cite{Rosati23}
\begin{equation}\label{PL}
I(E,x,t)=\sum_\alpha  \frac{f^{\text{2D},\circ}_{\alpha}(x,\textbf{q}=0,t) \,\tilde{\gamma}_\alpha \left(\tilde{\gamma}_\alpha+\Gamma\right)}{(E-E_\alpha)^2+(\tilde{\gamma}_\alpha+\Gamma)^2}.
\end{equation}
The PL is influenced by the temperature-dependent exciton-phonon scattering rate $\Gamma=\hbar/2\tau_p$   and  the radiative decay rate $\tilde{\gamma}=\gamma_\alpha\vert \psi_{\alpha, \textbf{q}=0}\vert^2$, which is proportional to the  $q=0$ component of the squared wavefunction in center-of-mass momentum  $\psi_{q_x}$. This is due to the conservation of momentum during the emission process \cite{Feierabend19b}.  Finally, for the case of the time-integrated PL, we consider an exciton lifetime $\tau=$175 ps according to recent experiments in LHs \cite{Beret22}.

\begin{figure}[t!]
    \centering	\includegraphics[width=\columnwidth]{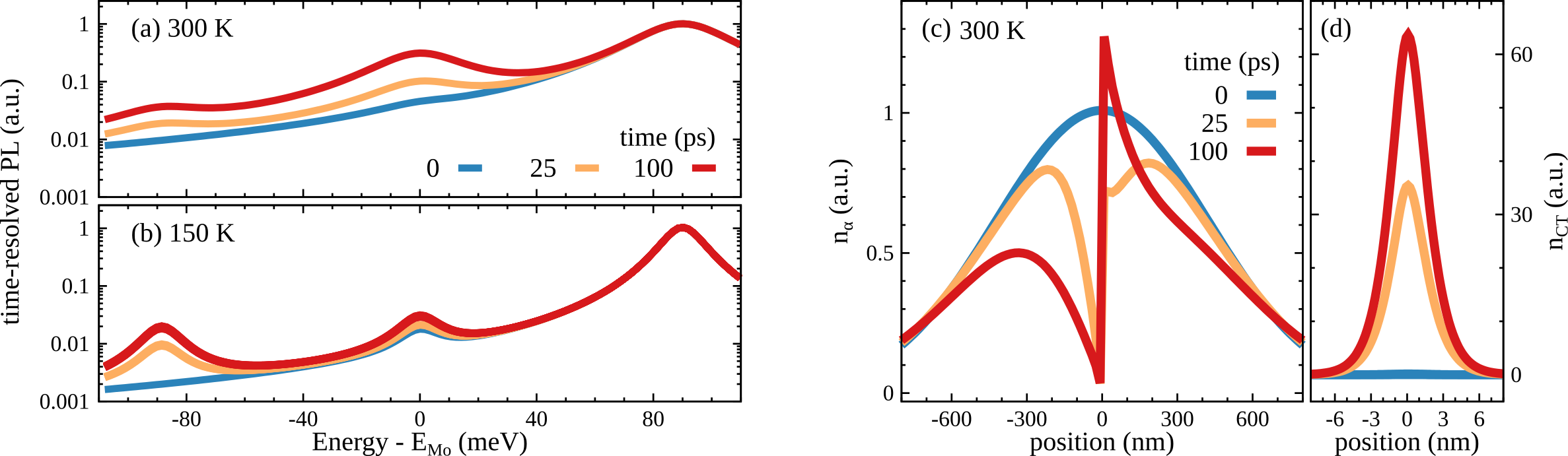}
	\caption{(a-b)Energy- and time-resolved photoluminescence (PL) spectrum normalized to the X$_{\text{W}}$ peak after micrometer excitation at the interface in resonance with X$_{\text{W}}$ at (a) 300K and (b) 150 K. At 300 K, the X$_{\text{Mo}}$ resonance becomes comparable to X$_{\text{W}}$ after approximately 100 ps reflecting the efficient exciton drift across the interface. Such a propagation is suppressed at the smaller temperature of 150 K, resulting in X$_{\text{W}}$ remaining much larger than X$_{\text{Mo}}$. Furthermore, a small X$_{\text{CT}}$ peak  emitted from CT excitons appear after few hundreds of picoseconds. (c-d) Space-resolved densities of (a) WSe$_2$ and MoSe$_2$ excitons as well as of  (b) CT excitons  at 300 K after an off-resonant optical excitation at the interface. Similarly to the case of a resonant excitation (Fig. 2 of the main manuscript) we find the appearance of exciton depletion and accumulation  at the WSe$_2$ and MoSe$_2$ side of the interface, respectively. At the same time, we observe the build-up of a large CT-exciton density that is about two orders of magnitude larger than the optically excited density.
	}
	\label{fig_2}
\end{figure}

 In  the main manuscript (Fig. 3(c)) we showed the time-resolved evolution of the energy- and time-resolved, space-integrated PL after resonant excitation at the interface at room temperature. In Fig. S\ref{fig_2}(b) we illustrate the PL at 150 K. For a direct comparison, in Fig. S\ref{fig_2}(a) we show again the results at 300 K, now taking cuts at given times. In this way, we clearly observe the delayed appearance of the X$_{\text{Mo}}$ resonance, which is formed due to exciton propagation across the interface  driven by the energy offset of the optically excited WSe$_2$ excitons. Thanks to this unidirectional propagation, the X$_{\text{Mo}}$ resonance becomes as intense as X$_{\text{W}}$ already after few hundreds of picoseconds. Furthermore, the time-resolved PL allows to resolve a small peak X$_{\text{CT}}$ emitted by CT excitons. 
The situation becomes significantly different at 150 K, cf. Fig. S\ref{fig_2}(b). Here, the trapping into CT excitons becomes so efficient that only few optically excited WSe$_2$ excitons are able to cross the interface. As a consequence, the X$_{\text{Mo}}$ resonance remains one order of magnitude smaller than the X$_{\text{W}}$ peak even after 150 ps. The height of X$_{\text{Mo}}$  increases only weakly compared to its initial value induced by the residual non-resonant excitation ($c_L=0.01$ in Eq. \ref{initial}). Furthermore, contrary to the room temperature case also the X$_{\text{CT}}$ resonance becomes as intense as X$_{\text{Mo}}$. While the transport across the interface is suppressed by the decreasing temperature, the trapping still takes place, resulting in the formation of the X$_{\text{CT}}$ resonance.

 In the main manuscript, we have also considered a far-field excitation (laser spot of 1 $\mu$m) resonant to the WSe$_2$ exciton energy. In Fig. S\ref{fig_2}(c-d), we show the same situation, but now after a high-energy excitation, resulting in an initial exciton density being spatially symmetric around the interface, cf. the cyan line in Fig. S\ref{fig_2}(c). Exploring the 
 spatiotemporal exciton dynamics,  we observe  (i) an accumulation of exciton density at the MoSe$_2$ side and (ii) a depletion area at the WSe$_2$ side of the interface, cf. Fig. S\ref{fig_2}(c). This is induced by the unidirectional propagation driven by the energy offset between the two sides.  Furthermore, we find the formation of (iii) a large CT exciton density, cf. Fig. S\ref{fig_2}(d). It becomes one (two) orders of magnitude larger than the initial exciton density after approximately 10 ps (100 ps).

